\documentclass{aa}%

\usepackage[varg]{txfonts}
\usepackage{amsmath}
\usepackage{amssymb}
\usepackage{graphicx}

\newcommand{\SAS}{{\sc SAS}}
\newcommand\TopStrut{\rule{0pt}{2.6ex}}          
\newcommand\XMM{XMM-\emph{Newton}}
\newcommand\Swift{\emph{Swift}}
\newcommand\ergx{\,erg\,s$^{-1}$\,cm$^{-2}$}
\newcommand{\CSS}{V808~Aur}

\begin{document}%

\title{X-ray and optical observations of four polars\thanks{Based on observations obtained with \XMM\ , an ESA science mission 
   with instruments and contributions directly funded by 
   ESA Member States and NASA},
   \thanks{ Based on observations collected at the Centro Astronómico Hispano Alemán (CAHA) at Calar Alto, operated jointly by the Max-Planck Institut für Astronomie and the Instituto de Astrofísica de Andalucía (CSIC).}}
\author{H.~Worpel\inst{1}, A.~D.~Schwope\inst{1}, T.~Granzer\inst{1}, K.~Reinsch\inst{2}, R.~Schwarz\inst{1},  I.~Traulsen\inst{1}}
\institute{Leibniz-Institut f\"ur Astrophysik Potsdam (AIP), An der Sternwarte 16, 14482 Potsdam, Germany \and
           Institut f\"ur Astrophysik, Friedrich-Hund-Platz 1, 37077 G\"ottingen, Germany
          }
\date{}

\abstract {}
   {To investigate the temporal and spectral behaviour of four polar cataclysmic variables from the infrared to X-ray regimes,
    refine our knowledge of the physical parameters of these systems at different accretion rates, and to search for a possible
    excess of soft X-ray photons.}
   {We obtained and analysed four \XMM\ X-ray observations of three of the sources, two of them discovered with the SDSS, one in the RASS.
   The X-ray data were complemented by optical photometric and spectroscopic observations and, for two sources, archival \Swift\ observations.}
   { \emph{SDSSJ032855.00+052254.2} was X-ray bright in two \XMM\ and two \Swift\ observations, and shows transitions from high
     and low accretion states on a timescale of a few months. The source shows no significant soft excess. We measured the magnetic
     field strength at the main accreting pole to be 39\,MG, the inclination to be $45^\circ\leq i \leq 77^\circ$, and we have
     refined the long-term ephemeris. \\
     \emph{SDSSJ133309.20+143706.9} was X-ray faint. We measured a faint phase X-ray flux and plasma temperature for this source, which seems
           to spend almost all of its time accreting at a low level. Its inclination is less than about $76^\circ$.\\
     \emph{1RXSJ173006.4+033813} was X-ray bright in the \XMM\ observation. Its spectrum contained a modest soft blackbody
           component, not luminous enough to be considered a significant soft excess. We inferred a magnetic field strength
           at the main accreting pole of 20 to 25\,MG, and that the inclination is less than $77^\circ$ and probably less than $63^\circ$.\\
     \emph{V808 Aur}, also known as CSS081231:J071126+440405, was X-ray faint in the \Swift\ observation but there is nonetheless strong evidence for
     bright and faint phases in X-rays and  perhaps in UV. Residual X-ray flux from the faint phase is difficult to explain by thermal emission
     from the white dwarf surface, or by accretion onto the second pole. We present a revised distance estimate of 250\,pc.}
   { The three systems we could study in detail appear to be normal polars, with luminosities and magnetic field strengths typical
     for this class of accreting binary. None of the four systems studied shows the strong soft excess thought commonplace in polars
     prior to the \XMM\ era.}

\keywords{ stars: individual: SDSSJ032855.00+052254.2 -- stars: individual: SDSSJ133309.20+143706.9 -- 
           stars: individual: 1RXSJ173006.4+033813 -- stars: individual: V808 Aur --
           stars: cataclysmic variables -- X-rays: stars}

\titlerunning{Observations of four polars }
\maketitle%

\section{Introduction }
\label{sec:intro}

Polars are cataclysmic variable (CV) systems in which the white dwarf primary possesses a magnetic field strong
enough to prevent the formation of an accretion disc. In these systems, material lost by the companion travels down magnetic field
lines directly to an accreting pole of the white dwarf. 

As the ionised gas spirals around the magnetic field lines it emits cyclotron radiation, which appears from infrared to untraviolet wavelengths at the
$\sim$10-230~MG field strengths in these systems. When the accreting material nears the white dwarf surface a shock forms and
bremsstrahlung is emitted isotropically at X-ray wavelengths. Some of the bremsstrahlung is intercepted by the white dwarf surface and
re-emitted as a blackbody-like feature in the extreme UV to soft X-ray range (e.g. \citealt{Cropper1990}). Some polars exhibit accretion at both magnetic poles.
For such systems, the pole more distant from the donor star typically has a stronger magnetic field but accretes less gas (e.g. \citealt{Warner1995}, see also
Table 2 of \citealt{FerrarioEtAl2015} and references therein).

Many polars show an excess of energy in soft X-rays, far more than can be produced by re-emission of bremsstrahlung. This excess was
initially hard to explain but is now attributed to coherent blobs of gas penetrating deeply into the white dwarf atmosphere, so that
their energy escapes the white dwarf as a luminous blackbody-like component (e.g. 
\citealt{KuijpersPringle1982, RamsayCropper2004}). The soft excess led to many discoveries of polars with the EINSTEIN, EXOSAT, ROSAT,
and EUVE satellites and was suspected to be present in all polars. By the end of the ROSAT mission, over sixty soft excess polars were known (e.g. \citealt{ThomasEtAl2000}).

Every polar discovered in the \XMM\ era, however, has lacked a large soft excess. Some of these new sources, such as
V808 Aurigae \citep{WorpelSchwope2015b}, are still bright enough in soft X-rays to have been detected in the ROSAT
All-Sky Survey (they would have yielded at least six photons in the 0.1-2.4\,keV band; \citealt{VogesEtAl2000}) but probably would have been too faint to be easily identified as polars. 
It is less clear why no new polars with the soft excess have been seen in \XMM\ data. Though this is a pointed rather than a survey instrument, it is capable of discovering them serendipitously
\citep{VogelEtAl2008, RamsayEtAl2009} and, while slewing, has a sensitivity comparable to ROSAT's \citep{SaxtonEtAl2008}. Furthermore, several polars discovered optically in, for instance,
the Sloan Digital Sky Survey (SDSS) have had follow-up \XMM\ observations but
have not shown a strong soft-excess. Our knowledge of the intrinsic shape of the spectral energy distribution of a typical polar is still clearly incomplete.

To further investigate this issue we analyse \XMM\  data of three polar cataclysmic variables to look for possible soft excesses in
these objects. The sources studied are summarised in Sect. \ref{sec:source_cat}. Additionally, we have optical photometric observations from the Catalina Real-Time
Transient Survey (CRTS, \citealt{DrakeEtAl2009}), the Sloan Digital Sky Survey (SDSS, \citealt{EisensteinEtAl2011}), the STELLA robotic telescope \citep{StrassmeierEtAl2004},
and optical spectroscopy from Calar Alto. One source, SDSS J032855+052254.2, has previously unpublished archival \Swift\ observations that we also present here.

As well as seeking a soft X-ray excess we aim to use the X-ray and optical data to find the rotation periods, system inclinations,
and magnetic field strengths. These determinations contribute to a larger census of polars. 

Finally, we present a previously unpublished \Swift\ observation of the eclipsing polar V808 Aurigae in an X-ray faint state. The system was previously studied at
high and intermediate luminosities \citep{SchwopeEtAl2015-CSS-eph,WorpelSchwope2015b}. This work completes the characterisation of its X-ray and ultraviolet emission in
all accretion states. We also present a revised distance estimate for this source.

\subsection{The sources}
\label{sec:source_cat}

\emph{SDSS J032855+052254.2}, hereafter J0328,  was identified as a polar in the SDSS. It is moderately bright optically, measured at $g$ magnitude 18 in SDSS and varying between
magnitudes 17 and 21 in CRTS. It shows evidence of cyclotron humps in its SDSS spectrum, and variability in its circular polarisation and H$\alpha$ line velocity. The orbital period was determined
to be $121.97\pm 0.25$ minutes by \cite{SzkodyEtAl2007}, who estimated a magnetic field strength
of 33\,MG at the accreting pole.

\emph{SDSS J133309.20+143706.9}, hereafter J1333, is an optically faint (about 18.5 g magnitude; \citealt{SzkodyEtAl2009}) source identified as a polar in SDSS. Its orbital period was
measured to be $132\pm6$ minutes from radial velocity measurements of the H$\alpha$ line \citep{SchmidtEtAl2008}. It shows optical
brightness variations of about one magnitude at this period \citep{SouthworthEtAl2015}.

\emph{1RXS J173006.4+033813}, hereafter J1730, was detected as a moderately bright (0.1 counts/s) X-ray source in the 
Rosat All-Sky Survey \citep{VogesEtAl1999}, and was also detected by \emph{Swift} at about half this brightness in early 2006 \citep{ShevchukEtAl2009},
but \Swift\ observations in May 2009 showed the source to have dimmed to undetectability \citep{BhaleraoEtAl2010}. 
A 17th magnitude variable optical counterpart was discovered by \cite{DenisenkoEtAl2009}. Optical photometry and spectroscopy performed
by \cite{BhaleraoEtAl2010} gives an orbital period of $120.2090\pm0.0013$ minutes, a magnetic field strength of $\sim$42 MG, and an upper limit to
the distance of 830~pc.

\emph{V808~Aurigae}, previously known as CSS081231:J071126+440405, is a polar discovered in the Catalina Sky Survey \citep{DrakeEtAl2009, DenisenkoKorotkiy2009, TempletonEtAl2009}. It is
an eclipsing system with a period of 117.18\,min, eclipse duration 7.218\,min, and an inclination of $79.3^\circ-83.7^\circ$ \citep{SchwopeEtAl2015-CSS-eph}.
The companion star is probably an M4.6 red dwarf, as suggested by the period-secondary relations given in \cite{Knigge2006}. It has optical magnitudes  $m_V\approx21$ and $m_B\approx22.4$
\citep{ThorneEtAl2010}, consistent with this identification, from which its distance was estimated to be 390~pc \citep{WorpelSchwope2015b}. The variable star designation
V808 Aur was assigned in December 2015 \citep{KazarovetsEtAl2015}.

The source light curves show bright and faint phases in both optical and X-rays, indicative of an accreting pole moving in and out of view as the white dwarf
primary rotates. The source was studied at high and intermediate accretion states by \cite{WorpelSchwope2015b}, who found that at high accretion rates the 
other magnetic pole accretes visibly in X-rays and optical. There is also a light curve dip, caused by the accretion stream passing in front of the
main emitting spot, seen at optical wavelengths \citep{KatyshevaShugarov2012} and in X-rays, but surprisingly not in the ultraviolet \citep{WorpelSchwope2015b}. In that
paper, the magnetic field strengths of the two accreting poles were measured to be 36 and 69 MG at the primary and secondary poles, and their positions are only $140^\circ$ apart on the surface of the
white dwarf. The location of the primary accretion spot was found to depend on accretion rate, moving by $\approx 20^\circ$ from a trailing to a leading
longitude between the intermediate and high states.

\section{Data reduction and analysis}
\label{sec:reduction}
\subsection{\XMM}

Four \XMM\ observations of three of the four sources are available, summarised in Table \ref{tab:xmm_obs_log}.
The EPIC cameras \citep{StruderEtAl2001, TurnerEtAl2001} observed in full frame mode with the thin filter in all observations, and the Optical Monitor (OM, \citealt{MasonEtAl2001}) was operated
in imaging and fast modes with the UVW1 and UVM2 filters, with effective wavelengths of 2910\AA\ and 2310\AA\ respectively \citep{KirschEtAl2004}
We have not used the Reflection Grating Spectrometer \citep{denHerderEtAl2001} data due to poor signal-to-noise.

\begin{table}
 
 \caption{ Observation log of the X-ray observations. The lengths of the \Swift\ and GTI-filtered \XMM\ EPIC-$pn$ XRT exposures are given.}
 \label{tab:xmm_obs_log}
 \setlength{\tabcolsep}{6pt}
 \begin{tabular}{llllr}
  Target & OBSID & Inst. & Date & Exp (s)\\
  \hline
  J0328\textsuperscript{*} &  0675230201 & XMM  & 2012-01-27 & 2629\\
  J0328                    &  0675230701 & XMM  & 2012-02-18 &13042 \\
  J0328                    & 00045622001 & Swift& 2012-03-20 &  387 \\
  J0328                    & 00045622002 & Swift& 2012-06-29 & 3120 \\
  J0328                    & 00045622003 & Swift& 2012-07-03 &  617 \\
  J0328                    & 00045622004 & Swift& 2012-07-16 & 1073 \\
  \hline
  J1333\textsuperscript{\textdagger} &  0675230601 & XMM & 2012-01-26 & --- \\
  J1333\textsuperscript{*} &  0675230501 & XMM & 2012-01-26  &10803 \\
  \hline
  J1730                    &  0675230301 & XMM & 2012-02-19  & 18251 \\
  \hline
  V808 Aur                 & 00031326001 & Swift& 2009-01-08  &  7391 \\
  \multicolumn{4}{l}{\textsuperscript{*}\footnotesize{These exposure times are much shorter than the on-target time}} \\ 
  \multicolumn{4}{l}{\footnotesize{due to high radiation- see Sections \ref{sec:J0328_XMM} and  \ref{sec:J1333_XMM}.}}\\
  \multicolumn{4}{l}{\footnotesize{The associated OM exposures are longer.}} \\
  \multicolumn{4}{l}{\textsuperscript{\textdagger}\footnotesize{This nominally 7\,ks observation failed and was repeated}}\\
  \multicolumn{4}{l}{\footnotesize{later on the same day.}}
 \end{tabular}
\end{table}

We reduced the raw data with the Science Analysis System (SAS), version 14.0.0. The EPIC-\emph{pn} and EPIC-MOS
data were processed with the standard \emph{epchain} and \emph{emchain} tasks to generate calibrated event lists, and \emph{epreject}
was run for EPIC-\emph{pn} data. All timing data were corrected to the Solar System barycenter with the SAS \emph{barycen} task. The
data from all three X-ray instruments were filtered to exclude photons with energies below 0.2\,keV and above 15.0\,keV. The OM fast mode
data were reduced with the \emph{omfchain} task, with a bin size of 300\,s.

Our source extraction regions are circles centered on the source, with precise locations determined with the
\SAS\ \emph{edetectchain} task. The best extraction radii were found with the \emph{eregionanalyse} task. These are in the range 14-36
arcseconds in radius. The background extraction regions are rectangles lying as near to the source as practical, located so that source and background regions suffer approximately
the same charge transfer inefficiencies from photon registration to charge readout. A circular area 2.5 arcseconds larger than the source extraction region was
excluded from the background region. We produced X-ray light curves using the \emph{epiclccorr} task.

We fit the spectra with version 12.9.0k of {\sc Xspec} \citep{Arnaud1996}, and fitting the data from all three X-ray instruments
simultaneously. To avoid giving too much weight to bins with few photons, we used Churazov weighting \citep{ChurazovEtAl1996}. We fitted the
data between 0.2 and 10~keV for the EPIC-$pn$ instrument, and between 0.2 and 8.0~keV for the MOS cameras.

\subsection{\Swift\ observations}

Four observations of J0328, and one of \CSS, were taken by \Swift, listed in Table \ref{tab:xmm_obs_log}. We reduced these observations with the
\emph{xrtpipeline} task and extracted source photon event lists from a 20 pixel (47.1$''$) circular region surrounding the source. We judged the
position of the source visually. The background region was a large nearby region containing no source. To maximise the photon numbers, we did not impose
any energy cuts. In none of these observations were there enough source photons to obtain a useful spectrum. The photon arrival times were corrected to
the Solar System barycenter using the \emph{barycorr} task in the FTOOLS package \citep{Blackburn1995}.

We did not use data from the
Burst Alert Telescope. J0328 was not visible in the UVOT data, but \CSS\ was clearly visible. For this source the UVOT data were reduced with
the standard imaging mode data analysis pipeline described in the UVOT Software Guide\footnote{\url{http://swift.gsfc.nasa.gov/analysis/UVOT_swguide_v2_2.pdf}}
to produce properly calibrated, flat fielded, and exposure-corrected images. For each subexposure of the image file, source detection was performed with
the \emph{uvotdetect} task.

\section{Results}
\label{sec:results}

\subsection{J0328}

\subsubsection{CRTS and STELLA photometry}
\label{sec:J0328_crts}

The CRTS database (DR2, \citealt{DrakeEtAl2009}) lists 349 photometric observations
of J0328 between MJD 53644.41 and 56592.30 (2005 Oct 01 to 2013 Oct 27). The
source varied between a minimum brightness of 20.82 and a maximum of 17.13
(unfiltered, i.e.~white-light photometry). Only a few measurements revealed a 
brightness fainter than 20 mag. A more typical faint level seems to occur 
around mag 19.5. The light curve in original time sequence is indicative of
the occurrence of high and low states with a brightness increase of about 1--1.5
magnitudes within an orbital cycle and long-term variability of similar amplitude.

We downloaded the CRTS data and transformed the MJD timings to Barycentric Julian Dates using
code developed by \cite{EastmanEtAl2010} and provided via the web-pages of the Ohio
State University\footnote{http://astroutils.astronomy.ohio-state.edu/time/utc2bjd.html}.  
A period search using the analysis of variance method \citep{Schwarzenberg-Czerny1989} revealed one pronounced periodicity close to the expected
value: $P = (7324.00 \pm 0.02)$\,s which will be regarded as the orbital period of
the binary. The new period agrees with our STELLA results (see below) and those of \cite{SzkodyEtAl2007} and is 
of sufficient accuracy to connect all observations
presented  in this paper without cycle count error. The accumulated uncertainty
over the eight years of photometric CRTS observations, which covers almost 35000
revolutions of the binary, is 0.1 phase units.

Time-resolved photometric observations with STELLA/WiFSIP \citep{StrassmeierEtAl2004} 
were obtained during three nights in February 2012 through $g$ and $r$
filters, respectively, always under stable photometric conditions. A log of
the observations is given in Table~\ref{t:stella}. 

The raw data were corrected for electronic bias and dark current 
and were flatfielded with pipeline scripts provided by the instrument
developers \citep{GranzerEtAl2001}.  

Differential photometry was performed with respect to a nearby 
star at RA(2000) = 52.23305 degrees, DEC(2000) = 5.35920, with $ugriz$
magnitudes of 17.44, 15.97, 15.36, 15.09, 14.97. 

\begin{table}
\caption{Log of STELLA photometric observations of J0328 in 2012}
\begin{tabular}{lllll}
\hline
Date & Obs interval & Filter & Exp & \# exp \\
      & $+2455000$d  &        & (s) & \\
\hline
Feb 21 & 979.35442 - 979.49616 & g & 60 & 126 \\
Feb 22 & 980.35512 - 980.49345 & g & 60 & 123 \\
Feb 23 & 981.41883 - 981.49139 & g & 60 & 65 \\[1ex]
\hline
\end{tabular}
\label{t:stella}
\end{table}

\begin{figure}
\resizebox{\width}{!}{
\includegraphics[angle=270, width=0.48\textwidth]{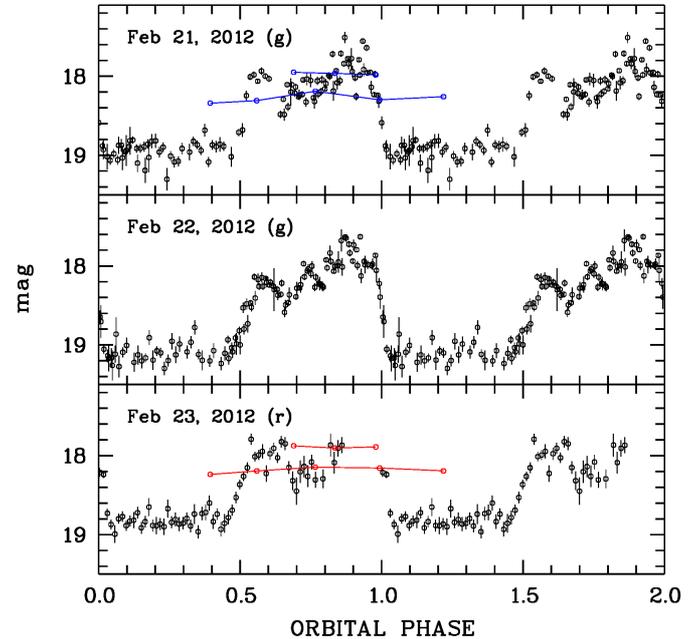}}
\caption{STELLA/WiFSIP photometry of J0328. The data were folded using the
  period of 122.1 min. Coloured symbols connected by lines indicate SDSS-spectrophotometry.}
\label{f:stella_j0328}
\end{figure}

The light curves in $g$- and $r$-bands display an on/off morphology 
similar to that reported by \cite{SzkodyEtAl2007} which was quite naturally 
explained due to the presence of an active pole. The amplitudes 
of the light curves were about 1.6 mag in the $g$-band and about 1.3 mag in
the $r$-band and are shown in Fig.~\ref{f:stella_j0328}. The rise to the 
bright phase is more gradual than the fall into the faint phase. There is
no indication of a second accretion region. The optical behaviour is thus
qualitatively similar to the X-ray (see Fig. \ref{fig:ltcrv_0328}), 
except for the lack of unambiguous evidence for a dip in the bright phase.

The end of the bright phase observed by STELLA on 2012 February 21 marks the zero point of the
long-term ephemeris. We have defined it to be the
time of half light of the decline to the faint phase, occurring at barycentric
Julian date BJD(TDB) $= 2455979.46865(6)$
in barycentric dynamical time.

The long-term ephemeris of J0328 thus becomes 
\begin{equation}
\mbox{BJD(TDB)} = 2455979.46865(6) + E \times 0.0847685(2).
\label{eqn:0328_eph}
\end{equation}


For this orbital period, the semi-empirical tables of \cite{Knigge2006} indicate
an M4.3 donor star of $0.177^{+0.023}_{-0.019}M_\odot$ assuming an uncertainty of 0.3 in the stellar type. That paper assumes a mean WD mass of 
$0.75 M_\odot$ with an intrinsic scatter of $0.16 M_\odot$, which we will adopt as the WD
mass and its uncertainty for this calculation. Since there are no eclipses, the inclination is
constrained to $i \lesssim 77^\circ$ using the method of \cite{ChananEtAl1976}. We observe the obscuration
of the accretion region by the accretion stream (see Section \ref{sec:J0328_XMM}), so the colatitude $\beta$ of the
magnetic pole must be less than $i$. We estimate from Fig. \ref{f:stella_j0328}
that the bright phase, when the accreting pole is in view, lasts for $\Delta \phi_B=0.55\pm 0.05$
of the binary orbit. From 
\begin{equation} 
\cot(i)=-\cos(2\pi\Delta\phi_B)\tan(\beta),
\end{equation}
we solve for $i_\text{min}=\beta$, assuming a spot with no vertical or lateral extent,
to find the lower limit to the inclination, giving
\begin{equation}
 i > \operatorname{arccot}\left[\sqrt{-\cos(2\pi\Delta\phi_B)}\right],
\end{equation}
or $i\geq 45.0^\circ$. 

\subsubsection{ \XMM\ X-ray observations}
\label{sec:J0328_XMM}

\begin{figure}
 \includegraphics{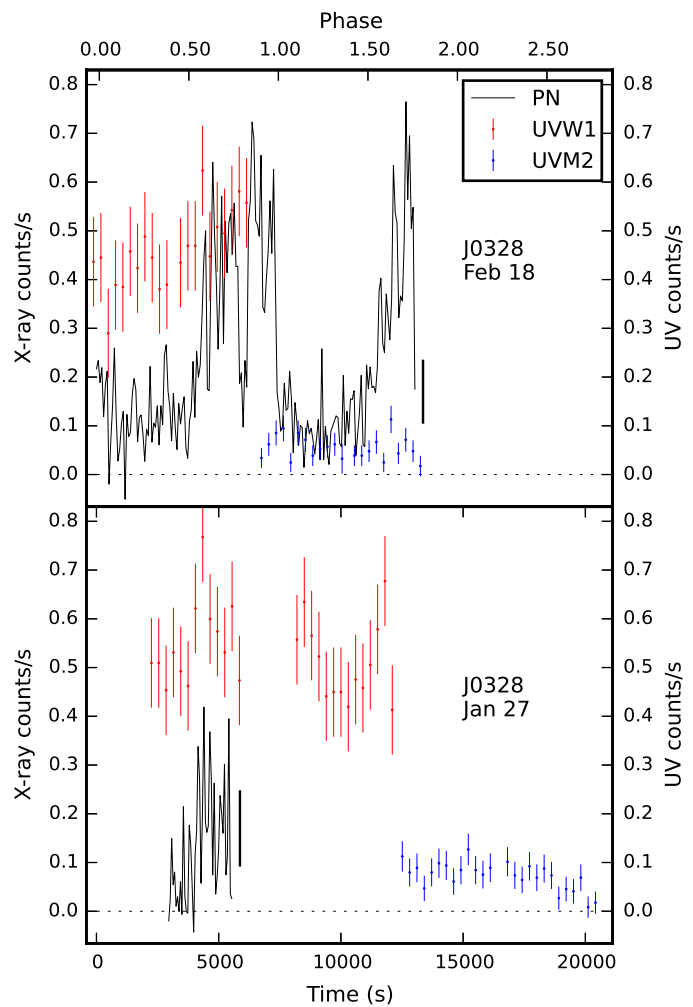}
 \caption{ EPIC-$pn$ and OM light curves of J0328 for the 2012 January 27 and February 18 observations, in the 0.2-10.0~keV energy band. The bin size is 60\,s for
           the X-rays and 300\,s for the OM. The shorter exposure has been offset to the correct orbital phase according to the ephemeris given in Eq. \ref{eqn:0328_eph}.
           Mean uncertainties on the X-ray light curve is indicated by the thick vertical line to its right.}
 \label{fig:ltcrv_0328}
\end{figure}

Two X-ray observations were made of this source, on 2012 January 27 and February 18. The light curves are shown in Fig. \ref{fig:ltcrv_0328}. 
The 2012 Feb 18 observation clearly shows distinct bright and faint phases characteristic of an emitting pole moving in and out of view.
The bright phase is asymetrically shaped, with a slow rise to a maximum brightness of 0.6 counts/s and a more rapid decline.
A deep dip in the bright phase is visible, lasting about seven minutes. This feature is likely due to the accretion stream passing in front
of the pole. The faint phase has residual X-ray emission of about 0.1 counts/s, possibly emission from a second accretion region, but because there
is no distinct second bump in the light curve this interpretation is uncertain. This observation is unaffected by proton flaring \citep{LumbEtAl2002},
except for a short period of about 0.2 counts per second at the beginning of the observation. It does not affect our analysis at all, since it occurs during the
faint phase.

According to the \XMM\ observation logs the 2012 Jan 27 observation was affected by radiation, causing the EPIC-$pn$ exposures to stop after 2,629~s of the
26\,ks on-target time, though the OM continued observing for considerably longer. Only around 470~s of data were recorded for the MOS instruments. In the remaining $pn$ data the source is faint ($\sim$ 0.1 counts/s)
but with an apparently rising intensity. According to our updated ephemeris (see Eq. \ref{eqn:0328_eph}), this observation occurred around the beginning of the
bright phase. As shown in Fig. \ref{fig:ltcrv_0328}, the source appears to have been slightly less luminous in both the bright and faint phases, perhaps
indicating a reduced accretion rate, but this observation is too short to be certain. There are not enough photons in this observation to obtain a useful
bright phase spectrum. The remaining X-ray data in this observation was completely unaffected by proton flaring.

\begin{figure}
 \includegraphics{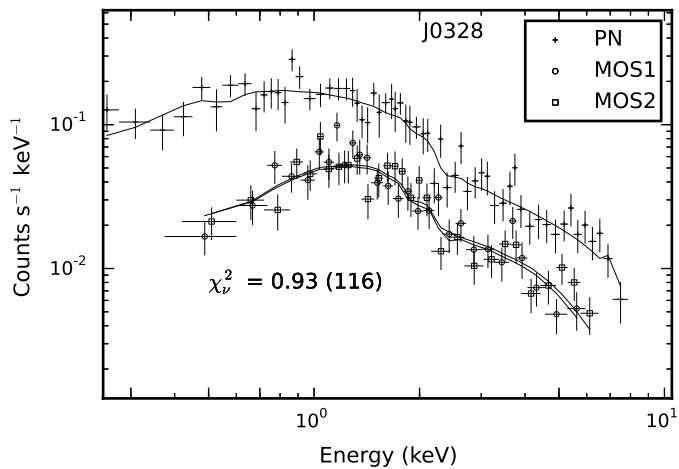}
 \caption{Bright phase X-ray spectrum of J0328. The data were fit for the three EPIC instruments simultaneously with a {\sc Mekal} plasma atmosphere model with both local and interstellar absorption.}
 \label{fig:spectralfits_0328}
\end{figure}

For the longer observation we extracted a bright phase spectrum, excluding the dip, for all three EPIC cameras. We fitted the spectra with a partially covered
{\sc Mekal} plasma atmosphere model \citep{MeweEtAl1985, LiedahlEtAl1995}, corrected for interstellar absorption with the \emph{phabs} component, i.e. {\tt 
phabs*pcfabs*mekal}. 
The plasma composition was left fixed at the Solar value. We calculated errors on variable parameters using the \emph{steppar} command, to 99\%
significance. The results of these fits are given in Table \ref{tab:spectralfits_0328}, and the spectrum is shown in Fig. \ref{fig:spectralfits_0328}.
For this source, the temperature of the plasma atmosphere was not strongly constrained. Since a pure plasma emission model suffices to describe the
data, there is no need to add any other radiation component.

\begin{table}
 \caption{ Spectral fits to the bright phase spectra of J0328. Normalisations
 for the {\sc Mekal} model are given in units of $10^{-17} \int n_e n_HdV / 4\pi[D_A(1+z)]^2$,
 where  $D_A$ is the angular diameter of the source, and $n_H,n_a$ are the hydrogen and
 electron densities. Normalisations for  the blackbody model are in units of $(R/D_{10})^2$
 where $R$ is the radius of the object in kilometers and $D_{10}$ is its  distance in units
 of 10~kpc. We give both absorbed and unabsorbed plasma atmosphere fluxes and in the last row
 we give the absorbed flux in the ROSAT band. Note that the temperature is fixed at 15\,keV.}
 \label{tab:spectralfits_0328}
 \begin{tabular}{ll}
  & J0328  \\
  \hline
  Interstellar $N_H$ ($10^{20}$~cm$^{-2}$) & $8.7^{+2.9}_{-1.9}$\TopStrut\\
  pcfabs $N_H$ ($10^{22}$~cm$^{-2}$) & $16^{+60}_{-11}$\TopStrut\\
  covering fraction & $0.27^{+0.05}_{-0.05}$\TopStrut\\
  $kT_\text{mekal}$ (keV) & $15$ \TopStrut \\
  norm$_\text{mekal}$ & $1.2^{+0.1}_{-0.1}$\TopStrut \\
  Flux$_\text{mekal,unabs}$ ($10^{-12}$ \ergx) &$3.1^{+0.2}_{-0.2}$ \TopStrut\\
  Flux$_\text{mekal,abs}$ ($10^{-12}$ \ergx) &$2.3^{+0.2}_{-0.1}$ \TopStrut\\
  $\chi^2_\nu$~(dof) & 0.95 (115)  \TopStrut\\
  Flux$_{(0.1-2.4),\text{abs}}$ ($10^{-13}$ \ergx) & $5.0^{+0.3}_{-0.3}$ \TopStrut\\
 \end{tabular}

\end{table}

\subsubsection{\XMM\ optical monitor observations}

The OM light curves are shown in Fig. \ref{fig:ltcrv_0328}. The source is faint and the quality
is poor, but in the UVW1 filter there are distinct indications of bright and faint phases corresponding
to those seen in X-rays. The source is detectable, though very faint, in UVM2 and shows no evidence of phase
modulation. The top panel clearly shows that the bright phase dip seen in X-rays is not present in the UV. 

It is conceivable that a large soft excess might be present in the extreme UV/very soft X-ray regime. We estimated the largest unobserved
blackbody in a manner very similar to the approach in \cite{WorpelSchwope2015b}. As reasoned in that paper, the unobserved soft excess can
come from a region no larger than the white dwarf surface (which we assume to be $R_{bb}<8000$\,km), its temperature can be no lower than that of the coolest known polar primary
(0.64 eV; \citealt{SchmidtEtAl2005, FerrarioEtAl2015}), its presence should not affect the observed X-ray spectrum, and it should not overpredict the UV flux arising from the accretion region in either
the UVW1 or UVM2 filters.

We simulated a series of blackbodies, with temperatures and normalisations on a grid with 0.5\,eV<$T_{bb}$<50\,eV and 1\,km<$R_{bb}$<8000\,km. The
blackbody effective radii are expressed as the radii of a sphere at the distance of J0328. The distance of J0328 is unknown, but we estimate it as follows.
We assume that at most 5\% of the flux of the system between 7440 and 7520\AA\ comes from the companion star. This scaling gives $m_R=19.44$.
Assuming a mass for the primary of $0.75M_\odot$, a spectral type for the companion of M4.0 to M4.6, and an inclination of $60^\circ$ (see Sect. \ref{sec:J0328_crts}) we
obtain effective radii for these stellar types by numerically computing the projected Roche lobe geometry for a range of orbital phases. 
Equations 7 and 8, and Table 2, of \cite{Beuermann2006} give a distance lower limit of 250\,pc.

This lower limit is less constraining than the recent work of \cite{CoppejansEtAl2016}, who give a minimum distance of 559\,pc. We point out that such a large distance appears unlikely because of J0328's
high Galactic latitude ($b=-40^\circ$). A distance of 559\,pc puts the source 360\,pc below the Galactic plane, or 1.6 -- 4.3 scale heights for the Galactic CV population \citep{RevnivtsevEtAl2008}.
We therefore adopted a conservative distance estimate of 500\,pc for the exercise of constraining the possible hidden soft excess. We then fitted the spectra between 200 and 1000\,eV with
the sum of the trial blackbody and the plasma atmosphere model of Table \ref{tab:spectralfits_0328}. Trials that caused the $\chi^2_\nu$ to increase by more than 1 were rejected. We
also rejected trials that overpredicted the accretion region flux density in either UV band, defined as the average bright phase flux density minus the average faint phase flux density,
plus the propagated $1\sigma$ uncertainty (see Fig. \ref{fig:ltcrv_0328}). The mean bright phase UV flux densities were formally $(6.7\pm6.8)\times 10^{-5}$\,keV\,cm$^{-2}$\,s$^{-1}$
and $(-0.6\pm2.7)\times 10^{-5}$\,keV\,cm$^{-2}$\,s$^{-1}$ for the UVW1 and UVM2 filters, i.e., both consistent with zero.

\begin{figure}
\includegraphics{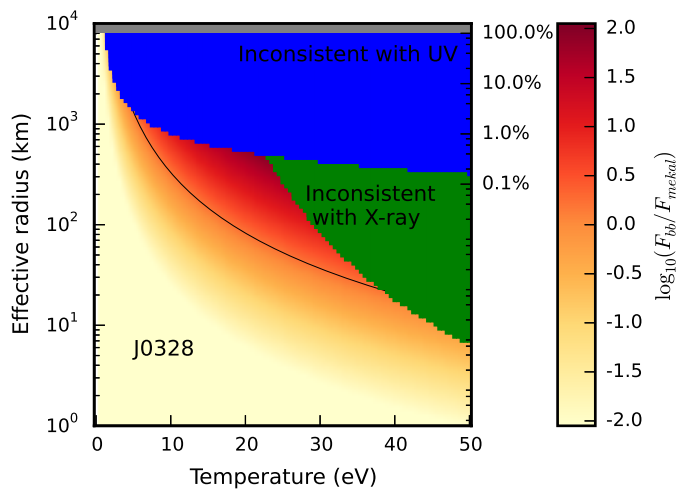}
\caption{Parameter space in which a soft excess can exist for J0328. The green regions indicate areas excluded because
 they conflict with the \XMM\ X-ray observations, the blue region indicates areas ruled out because they overpredict the UVW1 energy density,
 and the grey region indicates an emitting area larger than the surface of the assumed 8000~km radius WD. The shaded region indicates
 the magnitude of the potential soft excess (i.e. its flux divided by that of the X-ray plasma component). The black curve tracks where these are equal, indicating
 what would be expected for a bremsstrahlung being half intercepted and re-radiated. The right $y$-axis expresses the emitting area as a percentage of the WD surface.}
\label{fig:hiddenbb_0328}
\end{figure}

The bolometric flux of the potential soft blackbody was then compared to the flux of the plasma atmosphere model. The results are shown in Fig. \ref{fig:hiddenbb_0328}.
The largest soft excess formally compatible with the observations is 88 times as luminous as the plasma component, for a 22.5\,eV blackbody with an effective radius of 460\,km.
It is near to being ruled out by both the X-ray and UV data, and it is likely that more sensitive observations could have excluded it. Nevertheless it is possible that a substantial excess of flux can be
concealed in the very soft X-ray range. If J0328 were nearer or further than 500\,pc, then the boundaries of the green and blue regions would move down or up respectively.

\subsubsection{ \Swift\ X-ray observations }

\begin{figure}
 \includegraphics{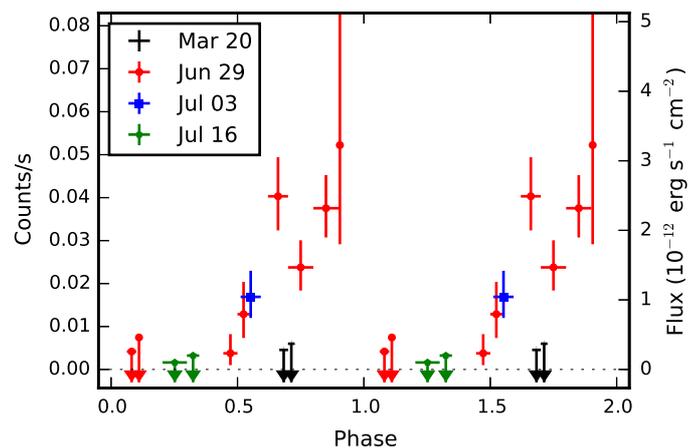}
 \caption{Phase-folded light curves of the four \Swift\ observations of J0328, obtained between
 March and July 2012. Two cycles are shown  for clarity. Points with no source counts have upper
 detection limits indicated by arrows. The right axis indicates the flux in the \XMM\ band assuming the bright
 phase spectrum listed in Table \ref{tab:spectralfits_0328}.}
 \label{fig:swift}
\end{figure}

The source was detectable only in the 2012 Jun 29 and July 3 observations, with a
total of 81 and 10 photons. Phase-folded light curves, according to the ephemeris in 
Eq. \ref{eqn:0328_eph}, are shown in Fig. \ref{fig:swift}. Though the phase coverage
is not complete, the bright and faint phases are evident in the longer observation.

The 2012 March 20 observation occurred during the bright phase. Its nondetection
suggests that J0328 was in a lower accretion state at that time. It is too early in phase to coincide with
the accretion dip. If we conservatively assume that we require six photons to detect the source, we get an
upper limit of 0.016 counts/s, less than half the brightness of the Jun 29 observation at the same phase.

The 2012 July 16 observation occurred during the faint phase, so the nondetection
of the source is unsurprising. The upper limit on the count rate is 0.0056 counts/s for 
six counts in a 1073\,s observation.

The short 2012 July 3 observation occurs during the rise to the bright phase. It well matches
the corresponding points in the longer Jun 29 observation, suggesting that the accretion rate did
not change much in the few days between those pointings.

We found the $1\sigma$ uncertainties on the background subtracted counts, following \cite{KraftEtAl1991}, by integrating over
the Poisson probability distributions, except that we also take into account the uncertainties in background rate as outlined in \cite{Helene1983}, Equation 5.

There are not enough photons in any \Swift\ observation to extract spectra but, assuming the same
spectral shape as the Feb 18 \XMM\ observation, a \Swift\ count rate of 0.01 counts per second
corresponds to a flux in the 0.2-10\,keV EPIC-$pn$ band of $6.2\times 10^{-13}$\ergx. This
suffices as a rough cross-calibration between the two sets of observations. The source has
returned to approximately its February 2012 brightness ($\sim2\times 10^{-12}$\ergx\ in
the bright phase) by late June.

\subsubsection{SDSS spectroscopy}
The SDSS \citep{EisensteinEtAl2011} observed J0328 spectroscopically two times using the same plate (plate number 2334). The
first observation happened through fibre 581 on MJD 53713, the second on MJD
53730 through fibre 600. These observations will henceforth be designated 2334-53713-581 and 2334-53730-600 (i.e. plate-mjd-fibre). 
The first observation consists of 5 sub-spectra of 15
min each covering binary cycles $-26728.6$ to $-26727.8$. The second observation 
had three sub-spectra covering binary cycle $-26529.3$ to $-26529.0$. 
Undulations of the continuum reminiscent of cyclotron harmonics were reported
by \cite{SzkodyEtAl2007} from follow-up spectroscopy of the source.

\begin{figure}
\includegraphics[angle=270, width=100mm]{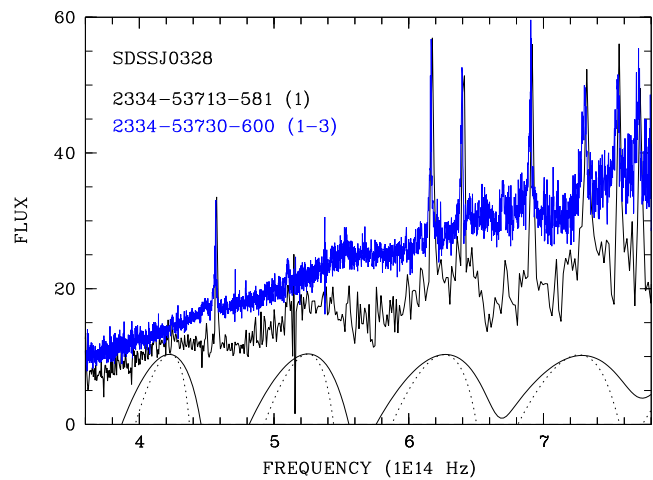}
\caption{SDSS spectra of J0328. Labels identify plate, MJD, fibre number and
  the sub-spectrum. The single 15 min spectrum obtained MJD53713 was median
  filtered for display purposes. Below the spectra cyclotron absorption
  coefficients originating from a 5 keV plasma are shown (see text for
  details). The units of flux are $10^{-17}$\ergx\AA$^{-1}$.}
\label{f:j0328_cyc}
\end{figure}

In order to gain further insight into the behavior
of the cyclotron features we inspected all the subspectra. It turned out that
the cyclotron features were found to be most obvious in the first sub-spectrum of 2334-53713-581.
This sub-spectrum and the yet unpublished average second spectrum are
shown in Fig.~\ref{f:j0328_cyc}. Cyclotron humps appear at about 
4.2, 5.2, and $6.3\times 10^{14}$\,Hz in 2334-53713-581
revealing a field strength of about
38\,MG. 

The observed spectra are shown together with the cyclotron
absorption coefficient for a 5 keV plasma at 38.5 MG (viewing angle $70\degr$) 
and 39 MG (viewing angle $80\degr$) which was normalized to a smooth
continuum beforehand. There is 1 MG uncertainty due to the unknown viewing
geometry, and another 1 MG uncertainty stems from the unknown plasma
temperature. Our revised field strength is  $39\pm2$\,MG. The field
strength of 33\,MG derived by \cite{SzkodyEtAl2007} identifies the hump 
at $5.2 \times 10^{14}$\,Hz with the 6th harmonic instead of the 5th 
and leads to a non-matching lower harmonic.

At the epoch of the second observation, 2334-53730-600, the object was brighter and the
continuum was less modulated by cyclotron emission. The three sub-spectra
cover the phase interval $0.7 - 1.0$ in cycle $-$26529. There appears to be 
a single hump at $5.5\times10^{14}$\,Hz. A single feature however is not
sufficient to determine a field strength. 

Synthetic SDSS-magnitudes were computed for each sub-spectrum by folding 
the spectral data through instrumental filter curves. The resulting light
curves are also shown in Fig.~\ref{f:stella_j0328}. On both occasions when
the bright phase was observed, the photometric variability was much lower
than expected from our own and published photometry \citep{SzkodyEtAl2007},
indicating a remarkable change in the accretion geometry of the source.

An estimate of the cyclotron flux of J0328 was derived by assuming that
the excess radiation forming the optical bright phase is pure cyclotron
radiation with a spectral shape given by the SDSS spectrum
2334-53730-600. This results in an observed integrated
cyclotron flux of $F_{\rm cyc} = 6.3 \times10^{-13}$\,\ergx.
The cyclotron spectrum peaks in the unobserved ultraviolet and is
expected to fall off rather quickly in the optically thin regime at even
shorter wavelengths. We estimate a bolometric correction factor of 1.3
which gives $F_{\rm cyc, bol} = 8.2 \times10^{-13}$\,\ergx\ which is lower
by a factor 3 than the plasma X-ray emission. Cyclotron cooling seems to
play a minor role in this high accretion state despite the rather large
field strength.

\subsection{J1333}

\subsubsection{\XMM\ observations}
\label{sec:J1333_XMM}
\begin{figure}
 \includegraphics{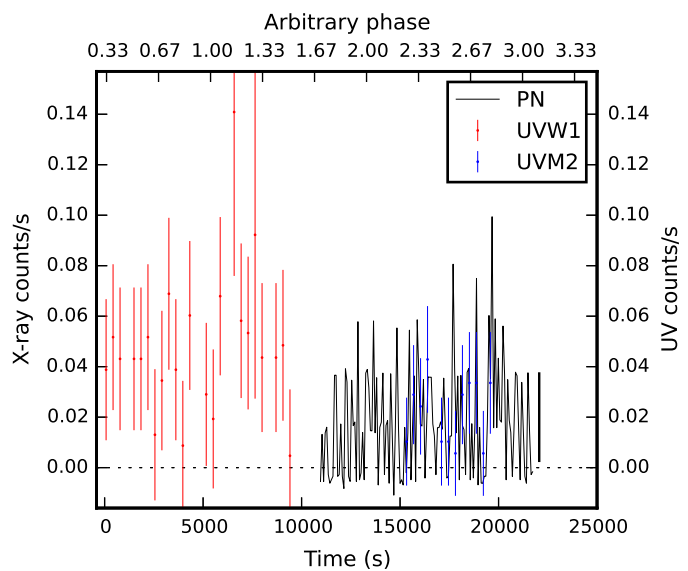}
 \caption{ EPIC-$pn$ and OM light curves of J1333 for the 2012 January 26 observation, in
           the 0.2-10.0~keV energy band. The bin size for X-rays is 60\,s, and 300\, for UV.
           Mean uncertainties on the X-ray data are indicated by the thick error bar to its right.} 
 \label{fig:ltcrv_1333}
\end{figure}

This source was observed by \XMM\ on 2012 Jan 26. Due to high radiation levels, \XMM\ only recorded for about 11~ks of the 22~ks on-target time,
but the remaining X-ray data in this observation were completely unaffected by proton flaring.
The X-ray light curve is shown
in Fig. \ref{fig:ltcrv_1333}. It is clear that the source was in a low accretion
state during the \XMM\ observation, and that distinguishing bright and faint phases
is impossible. The source was detectable using the SAS \emph{edetectchain} task,
and we extracted a very low resolution $pn$ X-ray spectrum. We fitted this spectrum
with a {\sc Mekal} model between 0.2 and 8~keV, giving a $\chi^2_\nu$ of 0.72 for
19 degrees of freedom, and the average flux over the observation was
2.75$\times 10^{-14}$\ergx. The inferred plasma temperature was low, about 4.3~keV.
The spectrum could be fitted just as well using a bremsstrahlung spectrum, or an
absorbed power law, but the measured flux in the 0.2 to 8~keV band did not change significantly
by changing the spectral model. The source was not bright enough for us to investigate any possible soft excess, or to
attempt to find any periodicity. 

\subsubsection{\XMM\ optical monitor observations}

The source was very faint (Fig. \ref{fig:ltcrv_1333}), with count rates of
$0.048\pm 0.028$ and $0.022\pm 0.013$ counts per second in the UVW1 and UVM2
bands. According to the conversion factors of \cite{KirschEtAl2004} these count
rates are equivalent to $(2.1\pm 1.2)\times 10^{-17}$\ergx\,\AA$^{-1}$
and $(4.7\pm 2.5)\times 10^{-17}$\ergx\,\AA$^{-1}$ for the UVW1 and
UVM2 filters respectively. There is no evidence for bright and faint phases.

\subsubsection{CRTS photometry}

\begin{figure}
 \includegraphics{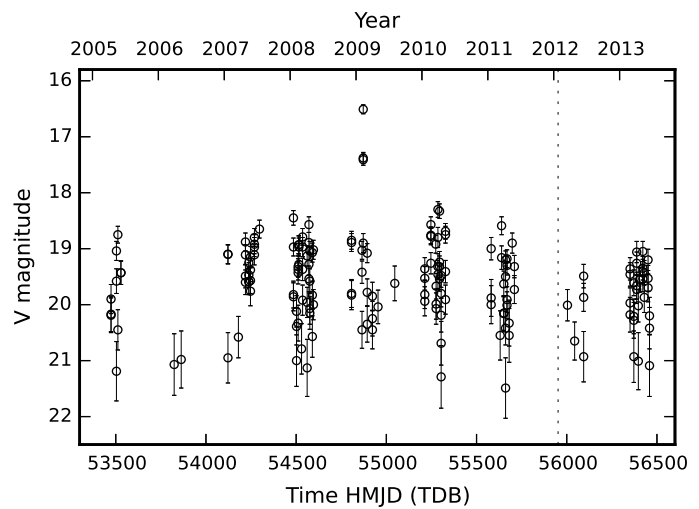}
 \caption{ CRTS light curve of J1333. The system remains between magnitudes 18.3 and 21.5,
           except for a brief brightening in early 2009. The time of the \XMM\ observation is
           indicated with a vertical dashed line.}
 \label{fig:J1333_crts}
\end{figure}

The CRTS database (DR2, \citealt{DrakeEtAl2009}) lists 171 photometric observations
of J1333 between MJD 53469.35 and 56454.26 (2005 April 9 to 2013 Jun 11). The
source varied between a minimum brightness of 21.49 and a maximum of 18.30
(unfiltered, i.e.~white-light photometry), except for a brief increase in brightness
to 16.51 on 2009 Feb 5 (see Fig. \ref{fig:J1333_crts}). We performed barycentric
corrections and a period search as in Sect. \ref{sec:J0328_crts}
around the inferred orbital period of 132~min \citep{SchmidtEtAl2008} but no
periodicities were detected.

There is no evidence for eclipses in our \XMM\ data, the CRTS light curves, or the light curves
presented in \cite{SchmidtEtAl2008, SouthworthEtAl2015}. The inclination is therefore less than
about $76^\circ$, following the arguments in Section \ref{sec:J0328_crts}. The orbital period of 2.2 hours
places this system in the orbital period gap, where the tables of \cite{Knigge2006} are incomplete,
but the companion masses either side of the period gap are about $0.2M_\odot$, slightly more massive than
the companion in J0328, so we have used $0.2M_\odot$ for the above calculation.

\subsection{J1730}

\subsubsection{\XMM\ observations}

\begin{figure}
 \includegraphics{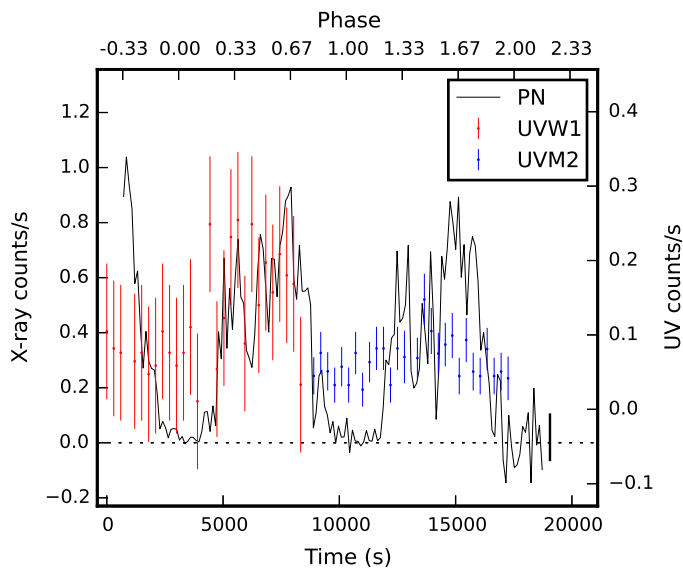}
 \caption{ EPIC-$pn$ and OM light curves of J1730 for the 2012 February 19 observation, in the
           0.2-10.0~keV energy band. The bin size is 120\,s for X-rays and 300\,s for UV.
           The mean uncertainty on the X-ray light curve is indicated by the thick error bar to
           the right. Phases are calculated according to the ephemeris of Equation \ref{eqn:1730_eph}.}
 \label{fig:ltcrv_1730}
\end{figure}

This source was observed by \XMM\ on 2012 February 19. The X-ray light curve is given
in Fig. \ref{fig:ltcrv_1730}. This light curve is typical of polars, showing a radiating
pole rotating in and out of view. Approximately 2.5 orbital periods were covered by \XMM\ and are shown in the Figure. 

The bright phases reach a peak brightness of $\sim 1$\,count\,s$^{-1}$. The first
half of the second cycle shows rapid, intense variability. There may be an accretion dip feature at around phase 0.33 but,
because of the variability in the light curve immediately before and after it, the evidence for a dip is not as clear as it was for J0328.
There is no residual X-ray emission in the faint phase, and hence no evidence for a second accretion region.

\begin{figure}
 \includegraphics{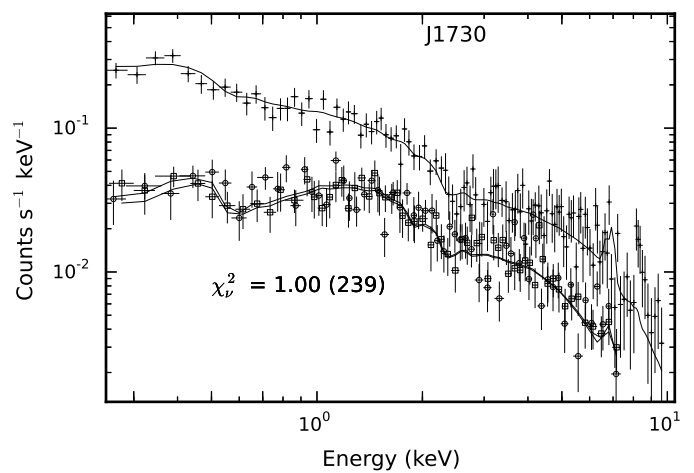}
 \caption{Bright phase X-ray spectrum of J1730, fitted for the three EPIC instruments simultaneously with a {\sc Mekal} plasma atmosphere
 model plus a large soft blackbody, corrected for both local and interstellar absorption.}
 \label{fig:spectralfits_1730}
\end{figure}

We extracted a bright phase spectrum for this source as we did for J0328. Fitting the J1730 data required some care. Using just a plasma atmosphere left large residuals at both the high and
low energy ends of the spectrum. We therefore had to use both a partially covering absorber to prevent implausibly high plasma temperatures, and a cool blackbody to account for an excess of soft photons.
We then modelled the spectra with the sum of a soft blackbody and a warmer Mekal plasma atmosphere, both absorbed by a partially covering photoelectric absorber and interstellar absorption: {\tt wabs*pcfabs*(bbodyrad+mekal)} in {\sc Xspec}. We imposed
an upper limit of $7.0\times10^{20}$\,cm$^{-1}$ for the interstellar absorption, the column density in the direction of J1730 \citep{DenisenkoEtAl2009}. We found that a slight excess of high energy
(>~ 5.0keV) photons causes the Mekal component to reach unphysically high temperatures. Thus, we fixed the plasma temperature at 15\,keV. 
The spectrum is shown in Fig. \ref{fig:spectralfits_1730}, and the fit parameters given in Table \ref{tab:spectralfits_1730}.
Changing the fixed plasma temperature to 10 or 20 keV does not significantly change the flux of this component, or of the blackbody.

\begin{table}
 \caption{ Spectral fits to the bright phase spectra of J1730. Normalisations
 for the {\sc Mekal} model are given in units of $10^{-17} \int n_e n_HdV / 4\pi[D_A(1+z)]^2$,
 where  $D_A$ is the angular diameter of the source, and $n_H,n_a$ are the hydrogen and
 electron densities. Normalisations for  the blackbody model are in units of $(R/D_{10})^2$
 where $R$ is the radius of the object in kilometers and $D_{10}$ is its  distance in units
 of 10~kpc. Spectral component fluxes are unabsorbed bolometric fluxes. In the last row we give the absorbed
 flux in the ROSAT band. Note that the plasma temperature is fixed at 15\,keV.}
 \label{tab:spectralfits_1730}
 \begin{tabular}{ll}
  & J1730  \\
  \hline
  Interstellar $N_H$ ($10^{20}$~cm$^{-2}$) & $5.0^{+2.0}_{-4.1}$\TopStrut\\
  pcfabs $N_H$ ($10^{22}$~cm$^{-2}$) & $6.4^{+7.1}_{-3.4}$\TopStrut\\
  covering fraction & $0.52^{+0.08}_{-0.09}$\TopStrut\\
  $kT_{bb}$ (eV) & $59^{+18}_{-12}$\TopStrut\\
  norm$_{bb}$    & 13,000$^{+86,000}_{-12,000}$\TopStrut\\
  $kT_\text{mekal}$ (keV) & $15$ \TopStrut \\
  norm$_\text{mekal}$ & $1.5^{+0.2}_{-0.2}$\TopStrut \\
  Flux$_\text{bb}$ ($10^{-12}$ \ergx) &$1.7^{+0.3}_{-0.3}$ \TopStrut\\
  Flux$_\text{mekal}$ ($10^{-12}$ \ergx) &$3.9^{+0.2}_{-0.2}$ \TopStrut\\
  $\chi^2_\nu$~(dof) & 1.00 (239)  \TopStrut\\
  Flux$_{(0.1-2.4)}$ ($10^{-13}$ \ergx) & $5.8^{+0.3}_{-0.2}$ \TopStrut\\
 \end{tabular}

\end{table}

We calculated the unabsorbed bolometric fluxes of the plasma and  blackbody components
between $10^{-6}$ and 100~keV using {\sc Xspec}'s \emph{cflux} tool. The chosen energy range
is wide enough that small changes to the upper and lower limits do not affect the result.
The component fluxes are also listed in the Table. The soft blackbody is two to three times less luminous than the plasma emission, and may
therefore represent part of the expected and sought-after reprocessed component but there is no evidence for a significant soft excess for this source either.

The field is not covered by CSS, hence no further long-term photometry
besides that from \cite{BhaleraoEtAl2010} is available to compare the new 
\XMM\ data with.

The bright phase for this source is longer than for J0328. The orbital period implies an M4.4 donor of $0.171 M_\odot$ and $0.208 R_\odot$.
We estimate $\Delta \phi_B\approx 0.6\pm 0.1$. Following the same reasoning as in Sect. \ref{sec:J0328_crts} 
the apparent absence of an accretion dip suggests that $i\lesssim 63^\circ$ because the inclination in this case must be less than the colatitude of the accretion spot.
However, the intense variability of the first half of the bright phase means that we cannot place much weight on this interpretation.
The lack of eclipses definitely constrains the inclination to $i \lesssim 77^\circ$, following the same procedure as in Section \ref{sec:J0328_crts}.

\subsubsection{\XMM\ optical monitor observations}
\label{sec:1730_xmmom}

The OM light curve is shown in Fig. \ref{fig:ltcrv_1730}. The source is faint, with a maximum
count rate of about 0.2\,count\,s$^{-1}$ for the UVW1 filter. It appears that the bright phase is visible in
the UVW1 data, and that it decreases to the faint phase before the X-rays do. It is not possible to tell
whether the bright phase dip visible in X-rays is present in UV, due to the faintness of the source.

The flux densities of the accreting pole emission, defined as the mean flux density of the bright phase minus that of the faint phase, were
$(2.1\pm2.4)\times 10^{-4}$\,keV\,cm$^{-2}$\,s$^{-1}$ and $(0.3\pm1.0)\times 10^{-4}$\,keV\,cm$^{-2}$\,s$^{-1}$ for the UVW1 and UVM2 filters.

\begin{figure}
  \includegraphics{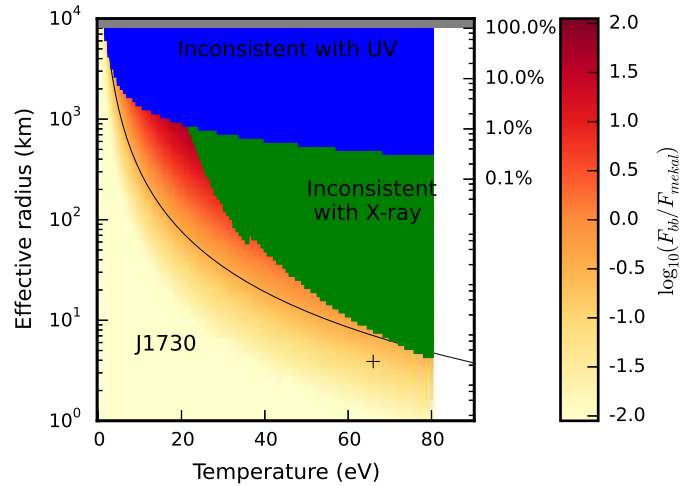}\\
   \caption{Parameter space in which a soft excess can exist for J1730. The green regions indicate areas excluded because
 they conflict with the \XMM\ X-ray observations, the blue region indicates areas ruled out because they overpredict the UVW1 energy density,
 and the grey region indicates an emitting area larger than the surface of the assumed 8000~km radius WD. The shaded region indicates
 the magnitude of the potential soft excess (i.e. its flux divided by that of the X-ray plasma component). The black curve tracks where these are equal, indicating
 what would be expected for a bremsstrahlung being half intercepted and re-radiated. The right $y$-axis expresses the emitting area as a percentage of the WD surface.
 The black cross shows the location of the soft blackbody we actually detected in this spectrum, and the diamond indicates the most luminous blackbody for
 a nearer (560\,pc) distance.}
  \label{fig:hiddenbb_1730}
\end{figure}

We estimated the largest possible soft blackbody flux in a similar manner to \ref{sec:J0328_XMM}, except that
the assumed distance is 830~pc and we substitute the grid soft blackbody for the one in the existing spectrum
rather than adding a completely new component. The results of this analysis are given in Fig. \ref{fig:hiddenbb_1730}. 
The maximum temperature is 80\,eV (roughly, the measured value plus the
upper uncertainty; see Table \ref{tab:spectralfits_1730}). For this source the largest possible unobserved blackbody ($T_{bb}=18.5$\,eV, $R_{bb}$=980\,km)
is 65 times as luminous as the plasma component, again for an emission region that covers a large area of the white dwarf surface and is only barely
consistent with the X-ray, UVW1, and UVM2 data. The 830\,pc distance is an upper limit; smaller distances would move the blue and green regions downward
and exclude these large soft blackbodies. For instance, applying Equation 3 of \cite{BhaleraoEtAl2010} to the companion radius predicted for J1730's orbital period in \cite{Knigge2006},
we obtain an estimated distance of 560\,pc. The most luminous hidden soft blackbody for this distance is indicated with a diamond on this Figure.
The discontinuity in the green curve is a point at which the flux at the low energy end is no longer underpredicted, but the $\chi^2_\nu$ is not too high for an acceptable fit.

\subsubsection{Calar Alto spectroscopy}

Low-resolution spectroscopy of J1730 was performed during the night of 2013
August 13 with the 2.2 m telescope of the Calar Alto observatory. The telescope was
equipped with CAFOS, a low-resolution grism spectrograph and imager. The G200
grism provided a wavelength coverage from 3750\,\AA\ up to 1.05 $\mu$m (useful
range below 9200\,\AA) at a FWHM
resolution of about 12\,\AA\ through a 1.2 arcsec wide slit, as 
measured from calibration lamp spectra. The integration time per spectrum was
5 m. A total of 18 spectra, covering slightly more than one orbital period, were taken.
The airmass varied from 1.23 to 1.65 over the series of exposures.

The spectrograph was rotated so that the light of the two stars 
USNO-B1.0 0936-00303765 (RA=17:30:02.78, DE=+03:38:32.8, $B=17.0$, $R=15.8$) 
and 
USNO-B1.0 0936-00303745 (RA=17:30:01.48, DE=+03:38:37.7, $B=16.1$, $R=15.5$) 
was also falling through the slit and could be used for the correction of
slit losses of the target star. 

Arc lamp spectra (Hg+He+Rb) for wavelength calibration and standard stars for 
photometric calibration were obtained before and after the sequence of the
target star. Standard star spectra were obtained through wide and narrow
slits. While the spectra that were obtained through wide slits were used to
determine the shape of the instrumental response curves, the one obtained through
the narrow slit was used for scaling purposes.

\begin{figure}
\includegraphics{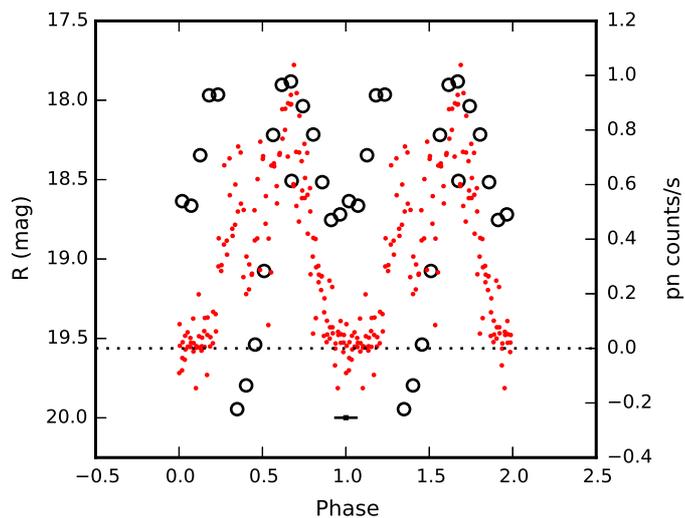}
\caption{R-band light curves derived from time-resolved spectrophotometric
  observations of J1730 on August 13, 2013 (large circles) and phase-folded EPIC-$pn$ count rates (small points).
  The cycle is plotted twice for clarity. The horizontal error bar indicates the maximum accumulated phase uncertainty
  between the optical and X-ray observations. The dotted horizontal line indicates zero $pn$ count rate.
}
\label{f:j1730_lcr}
\end{figure}

An approximate R-band light curve was extracted from the time-resolved
photometric spectra using the ESO-MIDAS software \citep{Warmels1992}, and is shown in Fig.~\ref{f:j1730_lcr}. The overall
brightness and the light curve shape indicate that J1730 was observed in an
active state, similar to epochs \#1 and \#4 of \cite{BhaleraoEtAl2010}. The large
amplitude modulation of the light curve is reminiscent of that shown by MR
Ser or V834 Cen in their high accretion states. In those sources the
optical light curves were modulated by strong cyclotron beaming.

We use the ephemeris defined by \cite{BhaleraoEtAl2010}:
\begin{equation}
 \text{BJD(TDB)} = 2454988.9375(3) + E\times 0.0834785(9).
 \label{eqn:1730_eph}
\end{equation} The \XMM\ and Calar Alto observations are 6483 cycles apart, and the uncertainty
in the period introduces an error of less than 0.006 in phase units. This is precise
enough to put the Calar Alto and \XMM\ light curves on a common scale.

\begin{figure}
\includegraphics[angle=270,width=95mm]{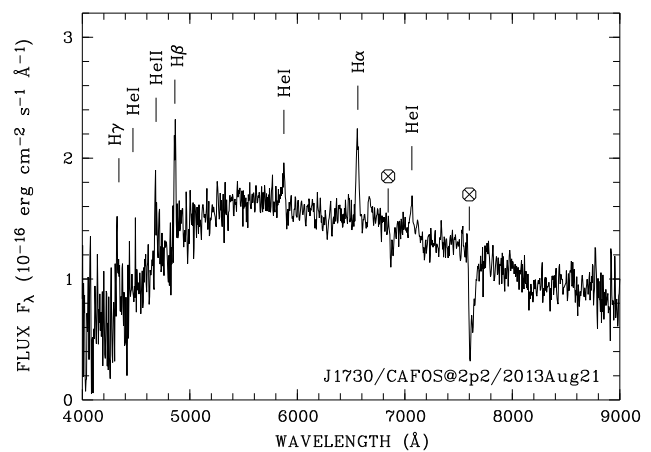}
\caption{Average mean bright spectrum of J1730 obtained Aug 21, 2013.
Main emission lines and uncorrected telluric absorption features
are labeled.}
\label{f:j1730spec}
\end{figure}

An average spectrum of the four brightest individual spectra is shown in Fig.~\ref{f:j1730spec}. 
Apart from H-Balmer, HeI and HeII emission lines that are typical for polars in their high accretion states 
the spectrum is dominated by a prominent continuum component which peaks at roughly 5500\,\AA.
It drops to $\sim$40\% of its peak value at 4000\,\AA. The decrease towards the red spectral 
regime is more moderate. This continuum spectral component is naturally explained as cyclotron radiation 
and the wavelength of the peak spectral flux indicates the turn-over of optically thick radiation at 
long wavelength to thin radiation at short wavelength/high harmonic number due to the strong 
frequency dependence of the cyclotron absorption coefficient. In a high state polar, the peak 
harmonic number is typically around 8 which gives a field estimate of around 25 MG. 
Objects with similar field strength in their main accretion regions are MR Ser and V834 Cen which 
show similar shapes of their high-state cyclotron continua \citep{SchwopeEtAl1993,SchwopeBeuermann1990}.
We note in particular that a field strength of 42 MG deduced by \cite{BhaleraoEtAl2010} cannot be confirmed 
by our observations. The data presented here would cover the harmonic range 3 -- 6 and one
would have most likely been able to resolve individual cyclotron harmonics in this low
harmonic regime. 

The integrated cyclotron flux at maximum phase is $7\times 10^{-13}$\,\ergx. The bolometric correction 
is an estimated 15\% so that the total flux at maximum phase becomes $8\times10^{-13}$\,\ergx.
The cyclotron flux is therefore fainter than the bremsstrahlung flux by a factor of around five (see
Table \ref{tab:spectralfits_1730}). Taking into account that the cyclotron emission is beamed and that the 
bremsstrahlung is re-radiated from the WD surface, their combined luminosity is
\begin{equation}
 L=\pi d^2\left(2 F_\text{brems} + \dfrac{3}{2}F_\text{cyc} + 2F_\text{bb}\right).
\end{equation}
From the upper distance limit of 830\,pc \citep{BhaleraoEtAl2010} we conclude that the total luminosity
is less than $2.6\times10^{32}$\,erg\,s$^{-1}$.

\subsection{V808 Aur (CSS081231:J071126+440405)}

\cite{WorpelSchwope2015b} provided a first estimate of the distance of 390\,kpc. A more sophisticated estimate using the method of
\cite{Beuermann2006} gives a nearer distance of about $250^{+50}_{-40}$\,pc.

\subsubsection{\Swift\ X-ray observations}
\label{sec:071126_swiftX}

The XRT \citep{BurrowsEtAl2005} detected \CSS\ emitting very weakly, with 26 source region photons over the 7.4\,ks observation, from a circular extraction region of 20 pixels (47.1$''$).
The background intensity was $4.5\times 10^{-3}$  counts per pixel, based on a large region containing no source, so there are probably five or six background photons in the source
extraction region. This equates to a source count rate of $2.8\pm 0.7\times 10^{-3}$ counts per second. There are clearly too few
photons to extract a spectrum, or to produce a meaningful light curve.

The 26 source region photons were divided into nominally bright and faint phase categories according to the ephemeris given in \citep{SchwopeEtAl2015-CSS-eph} and
assuming a leading spot with longitude $-11^\circ$, visible over 50\% of the orbit. There were 5 faint phase photons in 3,221 seconds 
of exposure, 20 bright phase photons in 3,440 seconds, and one during eclipse which must be background and is accordingly neglected. The count rates 
are therefore approximately $4.95\pm 1.3$ counts per kilosecond and $0.95\pm 0.7$ counts per kilosecond. Using Poisson counting statistics, we ruled out the possibility that these
count rates are equal at the 99.92\% level.

For assumed plasma temperatures of 5 to 15 \,keV the bolometric flux of the source during bright phase was about 2--5$\times 10^{-13}$\ergx. This is an order of magnitude lower than
the intermediate state ($3.86\times 10^{12}$\,erg\,s$^{-1}$\,cm$^{-2}$) observed in \cite{WorpelSchwope2015b} and implies an accretion rate of $\dot M=1$--$2\times 10^{-13} M_\odot \text{\,yr}^{-1}$ 
(see Eqn. 6.10 of \citealt{Warner1995}).


Some accretion emission must have been present during the faint phase, because the source was still visible then. An unabsorbed blackbody with the temperature of a typical WD primary
($\lesssim 15,000$\,K, e.g., \citealt{FerrarioEtAl2015}) would not have been detectable by \Swift\ over 3.2\,ks, as we verified with the \emph{fakeit} command of {\sc Xspec}. Nor
can the faint phase emission originate from the red dwarf donor star, for the same reason.

\subsubsection{\Swift\ UVOT observations}

The UVOT instrument \citep{RomingEtAl2005} used the UVW2 filter, which has a central wavelength of 1,928\AA. It was operated in image mode, so no photon arrival time information is available. 
The source was obviously visible in all five subexposures. The UVOT fluxes, as found with the \emph{uvotdetect} task, for the five subexposures are summarised in Table \ref{tab:uvot}. 
Each has been broken up into time contributions from the bright, faint, and eclipse phases. The mean X-ray photon barycenter correction of 454.9\,s was used to perform barycenter corrections
for the UVOT subexposure start and end times. 

\begin{table}
   \caption{The five UVOT sub-exposures. Fluxes in the UVOT-UVW2 band are given in units of $10^{-15}$\,erg\,s$^{-1}$\,cm$^{-2}$\,\AA$^{-1}$. For each measurement the uncertainty is
   $3\times10^{-17}$\,erg\,s$^{-1}$\,cm$^{-2}$\,\AA$^{-1}$. Times in seconds are given for the bright, faint,
   and eclipse phases, as well as their sum. }
   \begin{tabular}{llrrrrr}
   Exp & Flux & $t_\text{bri}$ & $t_\text{fai}$ & $t_\text{ecl}$ & $t_\text{tot}$ &Phase\\
   \hline
   1 & $1.33$ & ---    & 1493.2  & ---   & 1493.2  & 0.34-0.56\\
   2 & $1.29$ & 794.4  & 696.7   & ---   & 1491.1  & 0.17-0.38\\
   3 & $1.29$ & 1213.1 & ---     & 280.9 & 1494.0  & 0.99-0.20\\
   4 & $1.31$ & 985.6  & ---     & 433.1 & 1418.7  & 0.83-0.03\\
   5 & $1.47$ & 459.4  & 1034.9  & ---   & 1494.3  & 0.64-0.85\\
   \end{tabular}
   \label{tab:uvot}
\end{table}

The fluxes of the first four exposures are similar. However, the third and fourth exposures were partially during eclipse, when the flux should drop to almost zero,
whereas the first occurred entirely during the faint phase. This timing information suggests that the bright phase flux during Obs3 and Obs4 might be brighter than the Obs1
faint phase by $\sim 30-50\%$.

\section{Discussion}
\label{sec:discussion}

\subsection{J0328}

From multiple X-ray observations of J0328 we have found strong evidence that its accretion rate
varies strongly on timescales of a few months. It appeared to brighten slightly between Jan and
Feb 2012, before dimming significantly in March and brightening again by June and July.

We found an accretion dip for X-rays and visual light for J0328, but it was not present in UV. It is the
second source, after \CSS, to exhibit this phenomenon \citep{WorpelSchwope2015b}. A strong wavelength
dependence of the transparency of the accretion stream may provide clues regarding its structure
and physical properties. We have constrained the inclination of J0328 to approximately $45^\circ < i < 77^\circ$.

We have measured a magnetic field strength of 39$\pm 2$~MG for J0328 at the primary accreting pole,
based on the unique identification of cyclotron harmonics. This value is somewhat higher than the
33~MG previously measured by \cite{SzkodyEtAl2007}. 

If this source had been emitting with the same luminosity and spectral shape during the RASS,
it would have had a count rate of about 0.033 counts per second in the ROSAT 0.1-2.4\,keV band.
It would not have been bright enough to be listed
in the ROSAT Bright Sources Catalogue \citep{VogesEtAl1999}, but would probably have been detectable
in the Faint Sources Catalogue \citep{VogesEtAl2000}.

We investigated the possibility of a large unobserved soft excess in the very soft X-ray regime
and found that such a spectral component cannot be excluded but, if it exists, it can only be concealed
in a small area of the parameter space, with temperatures of less than about 25\,eV and an emitting area
of a few hundred kilometers in radius.

\subsection{J1333}

J1333 was in a faint state during the \XMM\ observation. It has never been observed in a
bright state in X-rays, but was discovered in the SDSS, where it displayed a prominent He\,II 4686\AA\ emission 
line caused by EUV photoexcitation. The CRTS photometry suggests that this system spends most of its time 
accreting at a low rate, with only brief and infrequent periods of high accretion. We obtained a mean 
X-ray flux of 2.75$\times 10^{-14}$\ergx\ and a plasma temperature of about 4\,keV, but were not able to detect any periodicity
in its light curve in either X-rays or UV. Further observations of this period-gap source would be helpful.
Its inclination is less than 76$^\circ$.

\subsection{J1730}

\begin{figure}
\includegraphics[angle=270,width=88mm]{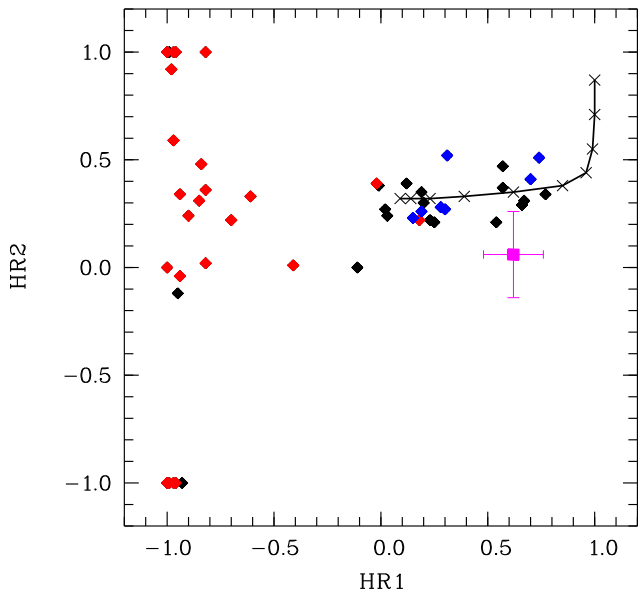}
\caption{X-ray colour-colour diagram of white dwarf accretors in the RBS (black
  -- nonmagnetic objects, blue -- intermediate polars, red --
  polars). The location of J1730 is shown with error bars. The black line
  delineates a thermal spectrum with 10 keV absorbed by cold interstellar
  matter. Log $N_{\rm H}$ was varied between 19 and 22 in steps of 0.33 dex, increasing from left to right. }
\label{f:rass_hrs}
\end{figure}

The object J1730 was detected in the ROSAT all-sky survey
(RASS) at an average count rate of $0.096 \pm 0.016$\,s$^{-1}$ during the 
total exposure of 472\,s. This is about twice as bright as in our \XMM\ observation, indicating that its X-ray luminosity can vary significantly.
The hardness ratios HR1 and HR2 of J1730 are compared
to other white dwarf accretors, mostly cataclysmic binaries, in the ROSAT
Bright Survey \cite[RBS, ][]{SchwopeEtAl2000} in Fig.~\ref{f:rass_hrs}. J1730 is
found in  a region occupied otherwise by non-magnetic CVs (dwarf novae and nova-likes)
or by intermediate polars. These sources display mainly thermal spectra and
the RASS X-ray colours of J1730 are indeed compatible with a thermal spectrum
with $kT > 5$\,keV absorbed by cold interstellar matter with $N_{\rm H} \simeq
2 \times 10^{20}$\,cm$^{-2}$. In particular they are lacking the soft
component shown by the majority of RASS-discovered or RASS-observed
polars. Only two among 23 RBS-polars showed hardness ratios compatible with 
thermal spectra, IW Eri and CD Ind (RBS541 and RBS1735), although with a
significantly lower absorbing column density. The underlying population will
be uncovered by upcoming sensitive surveys with e.g.~eROSITA \citep{MerloniEtAl2012}.

 The inclination of J1730 is not well constrained because there is no obvious accretion dip but, because of the lack of eclipses, it is definitely
less than $76^\circ$ and probably less than $63^\circ$.

\subsection{V808 Aur}

We have revised the estimated distance to this source downwards, to about 250\,pc. The analysis of the \CSS\ \Swift\ observation completes the X-ray and UV characterisation of the source begun in
\cite{WorpelSchwope2015b}. According to the light curve presented in Fig. 1 of \cite{SchwopeEtAl2015-CSS-eph} this observation occurred during the decline to a short low accretion state between two
high states. In X-rays the star was about an order of magnitude fainter than in an intermediate phase observed in 2012 by \XMM\ \citep{WorpelSchwope2015b}, but easily detectable in both wavelength regimes.

There is a strong ($\gtrsim 3.3\sigma$) indication of a phase-dependent brightness variation. The bright phase, as defined
by the ephemeris of the star and a sensible estimate of the accreting pole's location, is three times brighter than the faint phase. This result hints that an accreting pole is
active and distinct even at this low $\dot M$. The source is detectable in the faint phase, suggesting some luminosity beyond the thermal emission of the white dwarf itself. We could not
characterise the spectral differences between the bright and faint phases because not enough photons were detected. The bright phase in UV might be brighter than the faint phase by
$\sim 30-50$\% but the lack of timing information makes this difficult to determine.

The X-ray emission during the faint phase cannot be attributed to the companion star, or easily to a thermally emitting white dwarf surface. Accretion onto the distant hemisphere is a
possible explanation but this hypothesis presents difficulties. As pointed out by e.g., \cite{FerrarioEtAl1989}, transfer of material onto the distant pole prefers higher accretion rates.
For low $\dot M$ the accretion stream is less dense and is captured by magnetic field lines further from the primary. The field lines taking the gas to the distant pole often protrude outside
the Roche lobe of the primary, and it is energetically unfeasible for the gas to take this path. However, the wind-accreting pre-polars WX LMi \citep{SchwarzEtAl2001} and HS0922+1333 \citep{ReimersHagen2000}
show two-pole accretion geometry at very low accretion rate, demonstrating that a feasible trajectory to the second pole exists for some low accretion strongly magnetic CV systems. 

For the crude {\sc Mekal} approximation to the count rate developed in Sect. \ref{sec:071126_swiftX}, the ROSAT count rate would have been about 0.01 c/s. The sky position of \CSS\ was
viewed by ROSAT for a total of 192\,s in several scans separated by 96\, min. Even if it happened to observe the bright phase each time, this still would not have been enough for
\CSS\ to have been detectable even in the ROSAT faint source catalogue \citep{VogesEtAl2000}. The nondetection of \CSS\ in ROSAT is therefore consistent with an accretion state similar to this one.



\subsection{General remarks}
The dichotomy between polars discovered in the \XMM\ era, which all lack a prominent soft excess, and those identified as polars earlier, which all
have it, can still only partly be explained. Polars lacking this feature may have been detectable in the 0.1-2.4\,keV ROSAT band but would
appear as unremarkable faint sources, lacking distinguishing characteristics to motivate follow-up observations. This is probably true even
of polars such as \CSS\ and J1730 which require a modest addition of flux below 2.0\,keV to fit the \XMM\ spectra. Thus, their pre-\XMM\ non-discovery
is unsurprising.

The other question, why there have been no discoveries of soft excess polars in the \XMM\ era, remains unsolved. It cannot be purely due to the
lack of a post-RASS all sky X-ray survey because polars can be discovered by \XMM\ serendipitously \citep{VogelEtAl2008, RamsayEtAl2009} or in optical
surveys such as the SDSS. None of these have shown a soft excess. The upcoming eROSITA all-sky survey may uncover new objects of this kind, or provide
insights as to why they might be missing.

It may be that in these systems the missing soft excess is actually present but so cool that it is outside \XMM's detection band. Our work (this paper
and \citealt{WorpelSchwope2015b}) has consistently shown that a blackbody-like component, up to 50--100 times more luminous than the plasma component,
can be concealed in the extreme UV where it is invisible to \XMM's X-ray and optical telescopes. The current lack of a satellite sensitive in this energy
range makes testing this hypothesis difficult.


\begin{acknowledgements}
This work was supported by the German DLR under contract 50 OR 1405. This work has made use of the
Catalina Sky Survey. This study is based partly on data obtained with
the STELLA robotic telescope in Tenerife, an AIP facility operated jointly by AIP and IAC.
This research made use of Astropy\nocite{Astropy2013} and the PARI-GP computer algebra package.\nocite{PARI2}
\\
Funding for the Sloan Digital Sky Survey IV has been provided by
the Alfred P. Sloan Foundation, the U.S. Department of Energy Office of
Science, and the Participating Institutions. SDSS-IV acknowledges
support and resources from the Center for High-Performance Computing at
the University of Utah. The SDSS web site is www.sdss.org.
\\
SDSS-IV is managed by the Astrophysical Research Consortium for the 
Participating Institutions of the SDSS Collaboration including the 
Brazilian Participation Group, the Carnegie Institution for Science, 
Carnegie Mellon University, the Chilean Participation Group, the French Participation Group,
Harvard-Smithsonian Center for Astrophysics, 
Instituto de Astrof\'isica de Canarias, The Johns Hopkins University, 
Kavli Institute for the Physics and Mathematics of the Universe (IPMU) / 
University of Tokyo, Lawrence Berkeley National Laboratory, 
Leibniz Institut f\"ur Astrophysik Potsdam (AIP),  
Max-Planck-Institut f\"ur Astronomie (MPIA Heidelberg), 
Max-Planck-Institut f\"ur Astrophysik (MPA Garching), 
Max-Planck-Institut f\"ur Extraterrestrische Physik (MPE), 
National Astronomical Observatory of China, New Mexico State University, 
New York University, University of Notre Dame, 
Observat\'ario Nacional / MCTI, The Ohio State University, 
Pennsylvania State University, Shanghai Astronomical Observatory, 
United Kingdom Participation Group,
Universidad Nacional Aut\'onoma de M\'exico, University of Arizona, 
University of Colorado Boulder, University of Oxford, University of Portsmouth, 
University of Utah, University of Virginia, University of Washington, University of Wisconsin, 
Vanderbilt University, and Yale University.

\end{acknowledgements}

\bibliographystyle{aa}
\bibliography{bibli}

\end{document}